\documentclass[12pt]{article}
\usepackage{graphicx}
0.1cm \evensidemargin - 5cm
\newcommand{\eproof}{\rule{0.2cm}{0.2cm}}

\usepackage{latexsym,amssymb,amsmath,amscd,amsfonts}
\newtheorem{theorem}{Theorem}[section]
\newtheorem{proposition}[theorem]{Proposition}
\newtheorem{lemma}[theorem]{Lemma}
\newtheorem{corollary}[theorem]{Corollary}

\newtheorem{remark}[theorem]{Remark}

\begin{document}

\title{{{\bf Fokker-Planck-Kolmogorov equations associated with
SDEs driven by time-changed fractional Brownian motion}}}
\author{ Marjorie Hahn, Kei Kobayashi, Sabir Umarov }
\date{}
\maketitle
\begin{center}
{\it Department of Mathematics\\
Tufts University,
Medford, MA 02155, USA}\\


\end{center}


\begin{abstract}
In this paper Fokker-Planck-Kolmogorov type equations associated
with stochastic differential equations driven by a time-changed
fractional Brownian motion are derived. Two equivalent forms are
suggested.  The time-change process considered is either the first
hitting time process for a stable subordinator or a mixture of
stable subordinators. A family of operators arising in the
representation of the Fokker-Plank-Kolmogorov equations is shown to
have the semigroup property.
\end{abstract}



\section{Introduction}

In this paper we establish Fokker-Planck-Kolmogorov type equations
associated with stochastic differential equations driven by a
time-changed fractional Brownian motion. A (one-dimensional)
fractional Brownian motion $B_t^H$ is a zero-mean Gaussian process
with continuous paths and correlation coefficient
\begin{equation}
C_H(s,t)=E(B_s^H B_t^H)=\frac{1}{2} (s^{2H}+t^{2H}- |s-t|^{2H}),
\end{equation}
where the Hurst parameter $H$ takes values in $(0,1)$. 
If $H=1/2,$ then the correlation disappears, and $B_t^H$ becomes a
standard Brownian motion. Stochastic processes driven by a fBM are
of increasing interest for both theorists and applied researchers
due to their wide application in fields {such as} mathematical
finance, astrophysics, turbulence, etc.

Fractional Brownian motion $B_t^H,$ like standard Brownian motion,
has 
nowhere differentiable sample-paths. The covariance between
increments over non-overlapping intervals is positive, if 
{${1 \over 2}<H < 1,$} and negative, if $0<H<{1 \over 2}.$
Increments of $B_t^H$ exhibit long range dependence if ${1 \over
2}<H<1.$ The Hurst parameter $H$ can be extended to $H=1$ as well,
the corresponding fBM having the form $B_t^1=t N,$ where $N$ is the
standard normal random variable. If $H \in (0,1/2) \cup (1/2,1),$
then $B_t^H$ has the representation {\cite{Decr2,Nualart}}
$B_t^H=\int_0^t K_H(t,s)d B_s,$
where $K_H(t,s)$ 
is expressed through Gauss'  hypergeometric function.  FBM is not a
semimartingale unless $H=1/2$, so the usual {It\^o's} stochastic
calculus is not valid. Nevertheless, there are several approaches
\cite{Bender,book,Decr2,Nualart} to a stochastic calculus in order
to interpret in a meaningful way a SDE
driven by an $m$-dimensional fBM $B_t^{H}$ of the form 
\begin{equation} \label{sdefbm}
X_t=X_0+\int_0^t b(X_s) ds +\int_0^t \sigma(X_s)dB_s^H,
\end{equation}
where mappings $b: \mathbb{R}^n \to \mathbb{R}^n$ and $\sigma:
\mathbb{R}^n \to \mathbb{R}^{n \times m}$ are Lipschitz continuous
and bounded; $X_0$ is a random variable independent of $B_t^H.$ We
do not discuss in this paper these approaches  referring the
interested reader to \cite{book,CoutinDecr,Nualart}. {Instead, we
focus our attention on the} FPK equation associated with SDE
\eqref{sdefbm} driven by fBM whose form is
\cite{BaudoinCoutin,Gazanfur}
\begin{equation}
\label{KEfbm} \frac{\partial u(t,x)}{\partial t}= \sum_{j=1}^n
b_j(x) \frac{\partial u(t,x)}{\partial x_j}+ H t^{2H-1}
\sum_{j,k=1}^n a_{jk}(x) \frac{\partial^2 u(t,x)}{\partial x_j
\partial x_k},
\end{equation}
with the right hand side dependent on the time variable $t$, which,
in fact, reflects the presence of correlation. Functions $a_{jk}(x),
j,k=1,...,n$ are entries of the matrix $\mathcal{A}(x)=\sigma(x)
\times \sigma^T(x),$ where $\sigma^T(x)$ is the transpose of matrix
$\sigma(x).$ By definition $\mathcal{A}(x)$ is positive definite:
for any $x \in \mathbb{R}^n$ and $\xi \in \mathbb{R}^n$ one has
$\sum_{j,k=1}^n a_{jk}(x) \xi_j\xi_k \ge C |\xi|^2,$ where $C$ is a
positive constant. Additionally, $u(t,x)$ in equation (\ref{KEfbm})
satisfies the initial condition
\begin{equation} \label{icon}
u(0,x)= \varphi(x), ~ x \in \mathbb{R}^n,               
\end{equation}
where $\varphi (x)$ belongs to some function space, or is a
generalized function. In the particular case of FPK equation
associated with SDE \eqref{sdefbm}, $\varphi(x)=f_{X_{_0}}(x),$ the
density function of $X_0.$ If $X_0=x_0 \in \mathbb{R}^n,$ then
$\varphi(x)=\delta_{x_{_0}}(x),$ Dirac's delta with mass on $x_0.$
In this case the solution to {the} FPK equation is understood in the
weak sense.

In the sequel we use the following conventional notation:
\begin{equation} \label{oper1}
B(x,D_x)= \sum_{j=1}^n b_j(x) \frac{\partial}{\partial x_j}, ~~
A(x,D_x)= \sum_{j,k=1}^n a_{jk}(x) \frac{\partial^2 }{\partial x_j
\partial x_k},
\end{equation}
and
\begin{equation} \label{oper2}
L_{\gamma}(t,x,D_x)=B(x,D_x)+\frac{\gamma+1}{2}t^{\gamma}A(x,D_x),
\end{equation}
where $\gamma=2H-1.$ Due to the condition on the coefficients
$a_{jk}(x)$ stated above,  $A(x,D_x)$ is an elliptic operator. If
$\gamma=0,$ or equivalently $H=1/2,$ then the operator
$L_0(t,x,D_x)$ takes the form (coefficients not depending on $t$)
\begin{equation} \label{oper3}
L_0(t,x,D_x) \equiv L(x,D_x)=B(x,D_x)+ {1 \over 2} A(x,D_x),
\end{equation}
and equation (\ref{KEfbm}) coincides with the FPK equation
associated with the SDE driven by \linebreak Brownian motion $B_t$
(see, e.g. \cite{Stroock})
\begin{equation}
\label{KEbm} \frac{\partial u(t,x)}{\partial t}= L(x,D_x) u(t,x).
\end{equation}

The fractional FPK equation is obtained from equation \eqref{KEbm}
upon replacing the first order derivative on its  left hand side by
the time-fractional derivative $D_{\ast}^{\beta}$ in the sense of
Caputo-Djerbashian \cite{GM97}. By definition, the
Caputo-Djerbashian derivative of order $\beta$ is given by
\begin{equation}
\label{Caputo}
  {D}_{\ast}^{\beta} f(t) =
    {
      \frac{1}{\Gamma(1-\beta)}
      \int_0^{t} \frac{f^{\prime}(\tau) d \tau}{(t-\tau)^{\beta}}, ~
      0<\beta<1
    },
\end{equation}
where $\Gamma(\cdot)$ stands for Euler's gamma function. Introducing
the fractional integration operator
\[
J^{\alpha}f(t)=\frac{1}{\Gamma(\alpha)}\int_0^{t}
(t-\tau)^{\alpha-1}f(\tau)d\tau, ~~ \alpha >0,
\]
one can represent $D_{\ast}^{\beta}$ in the form
$D_{\ast}^{\beta}=J^{1-\beta}\frac{d}{dt}.$ We also write
$D_{\ast,t}^{\beta}$ emphasizing that the fractional derivative acts
with respect to the variable $t.$ An equivalent but slightly
different representation of the fractional FPK equation is possible
through the Riemann-Liouville derivative also, see e.g. \cite{Sok}.
The obtained Cauchy problem for the time-fractional FPK equation
\begin{align}
\label{fKEbm0} D_{\ast}^{\beta} v(t,x) &= L(x,D_x)v(t,x), t>0,~ x
\in
\mathbb{R}^n, \\
\label{0000} v(0,x) &= \varphi (x), ~ x \in \mathbb{R}^n,
\end{align}
describes the dynamics of a stochastic process driven by a
time-changed Brownian motion  (see Section \ref{Sec2}). Such
equations appear in many fields, including statistical physics
\cite{MetzlerKlafter00,Zaslavsky}, finance \cite{GMSR}, hydrology
\cite{BWM}, cell biology \cite{Saxton}, etc. 
 Existence and uniqueness theorems related to the Cauchy problem for
fractional differential equations, as well as more general
distributed order equations, can be found in 
\cite{EidelmanKochubey,Kochubey,ultra,UG05}. 
Instead, we focus on how fractional order FPK equations are obtained
from non-fractional FPK equations.

{By definition, a time-change process is a stochastic process with
continuous nondecreasing sample paths starting at $0$.} Let $B$ be a
standard Brownian motion {and $E$ be the time-change process given
by the first hitting time process for an independent stable
subordinator with index $\beta$.}
 If one replaces the driving process $B$ by a composition $B
\circ E,$ then the left hand side of equation \eqref{KEbm} becomes
the fractional derivative of order $\beta$, and the right hand side
remains unchanged. For details we refer the reader to
\cite{HKU,MWK}. As we will see, this is not the case for fractional
FPK equations associated with SDEs driven by time-changed fBM
(Section \ref{Sec3}).

Fractional FPK equations associated with SDEs driven by a
time-changed fBM (see equation \eqref{tchsdeFBM} below) have not yet
been determined. Meerschaert {et.}\ al.\ \cite{MNX} studied the 
{continuous time random walk (CTRW)} limits for {certain} correlated
random variables, which include linear fractional {L\'evy} stable
motions, and in particular, fractional Brownian motion. {For the
latter, the} scaling limits represent time-changed fBM, where the
time-change process is the inverse to a stable subordinator. Authors
of {that} paper {write, ``An} interesting open question is to
establish the governing equation for the CTRW scaling limit.''
A particular case of our Theorem 3.1 answers that question.

There are several approaches for deriving equation \eqref{fKEbm0},
including via semigroup theory \cite{MMMadded,HKU}, master equations
\cite{MMM-master,GM-master}, and continuous time random walks
\cite{GM1,GM2008,MMM-ctrw,US}. In this paper we use a different
technique, which can be extended for equations with a time dependent
right hand side as well, including equations of the form
\eqref{KEfbm}. This technique is close to the method used in
\cite{Kolakoltsev}.

SDEs driven by fBM are studied by several authors using different
approaches; for references we refer the reader to \cite{book}. SDEs
driven by time-changed Brownian motion are discussed in
\cite{Kobayashi}. The associated fractional FPK equations driven by
{time-changed} L\'evy processes when the {time-change} process is
the inverse to an arbitrary mixture of stable subordinators are
studied by Hahn, {et.}\ al.\ in \cite{HKU}. Note that any
time-changed semimartingale is again a semimartingale. However,
since fractional BM is not a semimartingale if $H \ne 1/2$, the
methods used in \cite{HKU} and \cite{Kobayashi} are not applicable
in this case. We plan to discuss a possible interpretation of SDEs
driven by a time-changed fractional Brownian and linear fractional
stable motion in a separate paper. Thus, in the present paper we
derive FPK type equations associated with the SDE
\begin{equation} \label{tchsdeFBM}
\displaystyle{X_t=x_0 + \int_0^t b(X_s)dE_s + \int_0^t
\sigma(X_s)dB_{E_{s}}^H},
\end{equation}
where $E_t$ is the inverse to an arbitrary mixture of stable
subordinators with indices 
{in $(0,1)$.} Throughout the paper we assume that $E_t$ is
independent of the driving process $B_t^{H}.$ An important
particular case {is} when $E_t$ is the inverse to a single stable
subordinator. The main ideas used in this paper will be illustrated
in this simpler case. The associated FPK equation can be represented
as a time-fractional order differential equation, but the right hand
side does not coincide with the right hand side of equation
\eqref{KEfbm}, unless $\gamma=0$ {(or equivalently, $H=1/2$)}.
However, in the case of zero drift (i.e.\ $b(x)\equiv 0$), the FPK
equation can be obtained with the same operator as on the right hand
side of (\ref{KEfbm}), but in this case the left hand side is
\textit{not} a time-fractional differential operator. This
difference of FPK equations is an essential consequence of the
correlation of the increments of the {fBM that is} the driving
process of the corresponding SDEs.

The paper is organized as follows. Section 2 illustrates the method
of this paper when the driving process is a time-changed Brownian
motion. The results obtained in this section further clarify
properties of density functions of processes which are inverses of
arbitrary mixtures of stable subordinators. In Section 3, {two equivalent} 
{FPK} equations associated with SDEs driven by time-changed fBM are
obtained extending the technique used in Section 2. {Furthermore,
the family of operators} appearing in the FPK equations is {shown to
have the semigroup property.}

\section{FPK equations associated with SDEs driven by a time-changed Brownian motion}
\label{Sec2}

Consider a SDE driven by a time-changed Brownian motion
\begin{equation} \label{tchsdeBM}
\displaystyle{X_t=x_0 + \int_0^tb(X_s)ds + \int_0^t
\sigma(X_s)dB_{E_{s}}}, ~ t>0,
\end{equation}
where $b(x)$ and $\sigma(x)$ are  Lipschitz continuous mappings and
$E_t$ is the first hitting time  process for a stable subordinator
$W_t$ with stability index ${\beta \in(0,1)}.$ The process $E_t$ is
also called an inverse to $W_t.$ The relation between $E_t$ and
$W_t$ can be expressed as $E_t= \min \{\tau: W_{\tau} \ge t\}.$ The
process $W_{t}, ~ t \ge 0,$ is a self-similar L\'evy process with
$W_0=0,$ that is $W_{ct}=c^{1 \over \beta} W_t$ as processes in the
sense of finite dimensional distributions, and its Laplace transform
is $\mathbb{E}(e^{-sW_t})=e^{-ts^{\beta}}.$ The density
$f_{W_{_1}}(\tau)$ of $W_1$ is infinitely differentiable on $(0,
\infty),$ with the following asymptotics at zero and infinity
\cite{MinardiLuchkoPagnini,UzhaykinZolotarev}:
\begin{align}
&f_{{W_{_1}}}(\tau) \sim \frac{({\frac \beta
\tau})^{\frac{2-\beta}{2(1-\beta)}}}{\sqrt{2\pi \beta (1-\beta)}} \,
e^{-(1-\beta)({\frac \tau  \beta})^{-\frac{\beta}{1-\beta}}}, \,
\tau
\to 0; \label{atzero}\\
&f_{W_{_1}}(\tau) \sim \frac{\beta}{\Gamma(1-\beta) \tau^{1+\beta}},
\, \tau \to \infty. \label{atinfinity}
\end{align} Since
$W_t$ is strictly increasing, its inverse process $E_t$ is
continuous and nondecreasing, but not a L\'evy process. Likewise the
time-changed process $B_{E_t}$ is also not a L\'evy process (see
details in \cite{HKU}). The associated FPK equation in this case has
the form
\begin{align}
\label{fKEbm} D_{\ast}^{\beta} v(t,x) &= L(x,D_x) v(t,x),
\end{align}
with the initial condition
$v(0,x)=\delta_{x_{_0}}(x),$
where $L(x,D_x)$ is defined in \eqref{oper3}, and $D_{\ast}^{\beta}$
is the fractional derivative in the sense of Caputo-Djerbashian.

Notice that solutions to equations \eqref{fKEbm} and \eqref{KEbm}
are connected by a certain relationship. Namely, a solution $v(t,x)$
to equation \eqref{fKEbm} satisfying the initial condition
\eqref{icon} can be represented through the solution $u(t,x)$ to
equation \eqref{KEbm}, satisfying the same initial condition
\eqref{icon}, by the formula
\begin{equation} \label{relation}
v(t,x)=\int_0^{\infty} f_t(\tau)u(\tau,x)d\tau,
\end{equation}
where $f_t(\tau)$ is the density function of $E_t$ for each fixed
$t>0.$ If $f_{W_{_1}}(t)$ is the density function of $W_1,$ then
\begin{equation} \label{relation2}
f_t(\tau)=-\frac{\partial}{\partial \tau}
Jf_{W_{_1}}(\frac{t}{\tau^{1 / \beta}}) = -\frac{\partial}{\partial
\tau} \int_0^{\frac{t}{\tau^{1 / \beta}}} f_{W_{_1}}(u)du
=\frac{t}{\beta \tau^{1+{1 \over \beta}}}f_{W_{_1}}(\frac{t}{\tau^{1
\over \beta}}).
 \end{equation}
Since $f_{W_{_1}}(u) \in C^{\infty}(0,\infty),$ it follows from
representation \eqref{relation2} that $f_t(\tau) \in
C^{\infty}(\mathbb{R}_+^2),$ where $\mathbb{R}_+^2=(0,\infty)\times
(0,\infty).$ Further properties of $f_t(\tau)$ are represented in
the following lemma.

\begin{lemma} \label{lemproperties}
Let $f_t(\tau)$ be the function given in \eqref{relation2}.  Then
\begin{enumerate}
\item[(a)] $\lim_{t\to +0}f_t(\tau)=\delta_0(\tau)$
~ in the sense of the topology of the space of tempered
distributions $\mathcal{D}^{\prime}(\mathbb{R});$
\item[(b)] $\lim_{\tau \to +0}f_t(\tau)=\frac{t^{-\beta}}{\Gamma(1-\beta)}, ~ t>0;$
\item[(c)] $\lim_{\tau \to \infty}f_t(\tau) = 0, ~ t > 0;$
\item[(d)] $\mathcal{L}_{t \to s}[f_t(\tau)](s)=s^{\beta-1}e^{-\tau s^{\beta}}, ~ s>0, ~ \tau \ge 0,$
\end{enumerate} \vspace{3mm}
where $\mathcal{L}_{t \to s}$ denotes the Laplace transform with
respect to the variable $t$.
\end{lemma}

\textit{Proof.} $(a)$ Let $\psi(\tau)$ 
 be an infinitely differentiable function rapidly decreasing at infinity.
We have to
show that $\lim_{t\to +0}<f_t,\psi>=\psi(0).$ Here $<f_t, \psi>$
denotes the value of $f_t \in \mathcal{D}^{\prime}(\mathbb{R})$ on
$\psi.$ We have
\begin{align*}
\lim_{t\to +0} <f_t(\tau),\psi(\tau)>&= \lim_{t\to
+0}\int_0^{\infty}f_t(\tau)\psi(\tau)d\tau \\
&= \lim_{t\to +0}\int_0^{\infty}f_{W_{_1}}(u)\psi \big((\frac{t}{u})^{\beta}\big)du\\
&=\psi(0) \int_0^{\infty}f_{W_{_1}}(u)du = \psi(0).
\end{align*}
Parts $(b)$ and $(c)$ follow from asymptotic relations
\eqref{atinfinity} and \eqref{atzero}, respectively. Part $(d)$ is
straightforward, just compute the Laplace transform of $f_t(\tau).$ \eproof

Due to part $(b)$ of Lemma \ref{lemproperties}, $f_t \in
C^{\infty}(0, \infty)$ for each fixed $\tau \ge 0.$ Hence, the
fractional derivative $D_{\ast,t}^{\beta}f_t(\tau)$ in {the}
variable $t$ is meaningful, and is a generalized function of
variable $\tau$.

\begin{lemma} \label{lemmain}
Function $f_t(\tau)$ defined in \eqref{relation2} for each $t>0$
satisfies the following equation
\begin{equation}
\label{relation3} D_{\ast,t}^{\beta}f_{t}(\tau)=
-\frac{\partial}{\partial \tau} f_t(\tau)
-\frac{t^{-\beta}}{\Gamma(1-\beta)}\delta_0 (\tau), 
\end{equation}
in the sense of tempered distributions.
\end{lemma}

\textit{Proof.} The Laplace transform (in variable $t$) of the left
hand side, using the definition \eqref{relation2} of $f_t(\tau),$
equals
$$\mathcal{L}_{t \to s}[D_{\ast,t}^{\beta}f_t(\tau)](s)=s^{\beta}
\mathcal{L}_{t \to s}[f_t(\tau)](s)-s^{\beta-1} \lim_{t \to +0}
f_t(\tau)= s^{2\beta -1}e^{-\tau s^{\beta}} -
s^{\beta-1}\delta_0(\tau), ~ s>0.$$ On the other hand, the Laplace
transform of the right hand side,
\begin{align*}
\mathcal{L}_{t \to s}{\Bigl[}-\frac{\partial}{\partial \tau}
f_t(\tau) -\frac{t^{-\beta}}{\Gamma(1-\beta)}\delta_0
(\tau){\Bigr]}(s) &=\frac{\partial^2}{\partial \tau^2}
(\frac{1}{s}e^{-\tau s^{\beta}})
- s^{\beta-1}\delta_0(\tau)\\
&= s^{2\beta -1}e^{-\tau s^{\beta}}-
s^{\beta-1}\delta_0(\tau), ~ s>0,
\end{align*} completing the proof.
 \eproof

\vspace{3mm}

\textbf{Derivation of fractional FPK equation.} Now it is easy to
show the derivation of the fractional order FPK equation
\eqref{fKEbm}, a solution of which is given by $v(t,x)$ in
\eqref{relation}. We have
\begin{align*}
D_{\ast,t}^{\beta}v(t,x)&=\int_0^{\infty} D_{\ast,t}^{\beta}
f_t{(\tau)}u(\tau,x)d\tau \\
& = -\int_0^{\infty}{\Bigl[}\frac{\partial}{\partial
\tau}f_t(\tau)+\frac{t^{-\beta}}{\Gamma(1-\beta)}\delta_0 (\tau){\Bigr]}  u(\tau,x) d \tau\\
  &= -\lim_{\tau \to \infty} [f_t(\tau)u(\tau,x)] + \lim_{\tau \to
  0}[f_t(\tau)u(\tau,x)] \\&+ \int_0^{\infty} f_t(\tau) \frac{\partial}{\partial
  \tau} u(\tau,x) d \tau -\frac{t^{-\beta}}{\Gamma(1-\beta)}u(0,x).
\end{align*}
Due to Lemma \ref{lemproperties}, part $(c)$ implies the first term
vanishes {since $u(\tau,x)$ is bounded}, while part $(b)$ implies
the second and last terms cancel. Taking into account \eqref{KEbm},
\begin{equation}
D_{\ast,t}^{\beta}v(t,x)= \int_0^{\infty} f_t(\tau) L(x,D_x)
u(\tau,x) d \tau = L(x,D_x) v(t,x).
\end{equation}
Moreover, by property $(a)$ of Lemma \ref{lemproperties},
$$\lim_{t \to
+0}v(t,x)=<\delta_0(\tau),u(\tau,x)>=u(0,x)=\delta_{x_{_0}}(x). \ \ \eproof$$

\vspace{3mm}

This technique extends to the more general case when the
{time-change} process is the first hitting time for an arbitrary
mixture of independent stable subordinators. Let $\rho(s)=\int_0^1
s^{\beta}d \mu (\beta),$ where $\mu$ is a finite measure {with $supp
\, \mu \subset (0,1].$} Let $W^{\mu}_t$ be a nonnegative stochastic
process satisfying $\mathbb{E}(e^{-sW^{\mu}_t})=e^{-t \rho(s)},$ and
$E^{\mu}_t=\min\{\tau: W^{\mu}_{\tau} \ge t\}.$ The process
$W_t^{\mu}$ represents a mixture of independent stable subordinators
with a mixing measure $\mu$ (see \cite{HKU}).

\begin{theorem} \label{Thdode}
Let $u(t,x)$ be a solution of the Cauchy problem
\begin{align} \label{fpkdode}
\frac{\partial u(t,x)}{\partial t} &= L(x,D_x) u(t,x), ~ t>0, ~ x
\in \mathbb{R}^n, \\
u(0,x)&=\varphi(x), ~ x \in \mathbb{R}^n. \label{fpkdodein}
\end{align}
 Then the function
$v(t,x)=\int_0^{\infty} f^{\mu}_t(\tau) u(\tau,x),$ where
$f^{\mu}_t(\tau)$ is the density function of $E^{\mu}_t,$ satisfies
the Cauchy problem
\begin{align}
D_{\mu}v(t,x) &\equiv \int_0^1 D_{\ast,t}^{\beta} v(t,x)d \mu
{(\beta)}= L(x,D_x) v(t,x), ~ t>0, ~ x
\in \mathbb{R}^n, \\
v(0,x)&=\varphi(x), ~ x \in \mathbb{R}^n.
\end{align}
\end{theorem}

The proof of this theorem requires two lemmas which generalize
{Lemmas} \ref{lemproperties} and \ref{lemmain}. Define the function
\begin{equation} \label{phi}
\Phi_{\mu}(t)=\int_0^{1}\frac{t^{-\beta}}{\Gamma(1-\beta)}d\mu(\beta),
~ t>0.
\end{equation}

\begin{lemma} \label{lemgen}
Let $f^{\mu}_t(\tau)$ be the function  defined in Theorem
\ref{Thdode}. Then
\begin{enumerate}
\item[(a)] $\lim_{t\to +0}f^{\mu}_t(\tau)=\delta_0(\tau), ~ \tau \ge 0;$
\item[(b)] $\lim_{\tau \to +0}f^{\mu}_t(\tau)= \Phi_{\mu}(t), 
~ t>0;$
\item[(c)] $\lim_{\tau \to \infty}f^{\mu}_t(\tau) = 0, ~ t \ge 0;$
\item[(d)] $\mathcal{L}_{t \to s}[f^{\mu}_t(\tau)](s)= \frac{\rho(s)}{s} e^{-\tau \rho(s)}, ~ s>0, ~ \tau \ge 0.$
\end{enumerate}
\end{lemma}

\textit{Proof.} First notice that
$f^{\mu}_t(\tau)=f_{E^{\mu}_t}(\tau)=-\frac{\partial}{\partial
\tau}Jf_{W^{\mu}_\tau}(t),$  where $J$ is the usual integration
operator. The proofs of parts $(a)-(c)$ are similar to the proofs of
parts $(a)-(c)$ of Lemma \ref{lemproperties}. Further, using the
definition of $W^{\mu}_t,$
\begin{equation*}
\mathcal{L}_{t \to s}[f^{\mu}_t(\tau)](s)= -\frac{1}{s}
\frac{\partial}{\partial \tau} \mathcal{L}_{t \to
s}[{f_{W_\tau^{\mu}}(t)}](s) =\frac{\rho(s)}{s} e^{-\tau \rho(s)}, ~
s>0,
\end{equation*}
which completes the proof. \eproof

\begin{lemma} \label{lemmaingen}
The function $f^{\mu}_t(\tau)$ defined in Theorem \ref{Thdode}
satisfies for each \, $t>0$ the following equation
\begin{equation}
\label{relation3} D_{\mu,t}f^{\mu}_{t}(\tau)=
-\frac{\partial}{\partial \tau} f^{\mu}_t(\tau) -\delta_0 (\tau)
\Phi_{\mu}(t), 
\end{equation}
in the sense of tempered distributions.
\end{lemma}

\textit{Proof.} Integrating both sides of the equation
$\mathcal{L}_{t \to s}\big[D_{\ast,t}^{\beta}f^{\mu}_{t}(\tau)\big]
= s^{\beta} \mathcal{L}_{t \to s} [f^{\mu}_{t}(\tau)]{(s)} -
s^{\beta-1} \delta_0(\tau), $ and taking into account part $(d)$ of
Lemma \ref{lemgen}, yields
$$
\mathcal{L}_{t \to s}\big[D_{\mu,t}f^{\mu}_{t}(\tau)\big] =
\frac{\rho^2(s)}{s} e^{-\tau \rho(s)} - \frac{\rho(s)}{s}
\delta_0(\tau).
$$
It is easy to verify that the latter coincides with the Laplace
transform of the right hand side of \eqref{relation3}.  \eproof

\vspace{3mm}

\textit{Proof of Theorem \ref{Thdode}.} Using Lemma
\ref{lemmaingen}, we have
\begin{align*}
D_{\mu,t}v(t,x)&=\int_0^{\infty} D_{\mu,t}
{f^\mu_t}{(\tau)}u(\tau,x)d\tau
  = -\lim_{\tau \to \infty} [f^{\mu}_t(\tau)u(\tau,x)] + \lim_{\tau \to
  0}[f^{\mu}_t(\tau)u(\tau,x)] \\&+ \int_0^{\infty} f^{\mu}_t(\tau) \frac{\partial}{\partial
  \tau} u(\tau,x) d \tau -\Phi_{\mu}(t) u(0,x) = \int_0^{\infty} f^{\mu}_t(\tau) \frac{\partial}{\partial
  \tau} u(\tau,x) d \tau,
\end{align*}
since all the limit expressions vanish due to {parts} $(b)$ and
$(c)$ of Lemma \ref{lemgen}. Now taking into account equation
\eqref{fpkdode},
\begin{equation}
D_{\mu,t}^{\beta}v(t,x)= \int_0^{\infty} f^{\mu}_t(\tau) L(x,D_x)
u(\tau,x) d \tau = L(x,D_x) v(t,x).
\end{equation}
The initial condition \eqref{fpkdodein} is also verified by using
property $(a)$ of Lemma \ref{lemgen}: 
$$\lim_{t \to
+0}v(t,x)=<\delta_0(\tau),u(\tau,x)>=u(0,x)=\varphi(x). \ \ \eproof$$

\section{FPK equations associated with SDEs driven by time-changed fBM}
\label{Sec3}

Now let us focus on the FPK equation associated with the
{SDE \eqref{tchsdeFBM}} driven by a time-changed fBM {$B^H_{E_t}.$} 
Recall that the FPK equation associated with a SDE driven by a fBM
(without time-change) has the form \cite{BaudoinCoutin,Gazanfur}
\begin{equation}
\label{KEfbm1} \frac{\partial u(t,x)}{\partial t}=
L_{\gamma}(t,x,D_x) u (t,x),
\end{equation}
where $L_{\gamma}(t,x,D_x)$ is defined in \eqref{oper2}
 and the Hurst parameter $H$  is connected with
$\gamma$ via $2H-1=\gamma.$  Again for simplicity, we consider a
time-change process $E_t$ inverse to a single stable subordinator
$W_t,$ though mixtures of stable subordinator can be treated
similarly. Hence, the density function $f_t(\tau)$ of $E_t$
possesses all the properties mentioned in Lemmas \ref{lemproperties}
and \ref{lemmain}.

\begin{theorem} \label{thfbm}
Let 
$u(t,x)$ be a solution to the Cauchy problem
\begin{align}
\label{CauchyFBM} \frac{\partial u(t,x)}{\partial t}&=
B(x,D_x)u(t,x)+\frac{\gamma
+ 1}{2} t^{\gamma} A(x,D_x)  u (t,x), ~ t>0, ~ x \in \mathbb{R}^n,\\
{u}(0,x)&=\varphi(x), ~ x \in \mathbb{R}^n. \label{incon}
\end{align}
Let $f_t(\tau)$ be the density function of the process inverse to a
stable subordinator of index $\beta.$ Then
$v(t,x)=\int_0^{\infty}f_t(\tau)u(\tau,x)d\tau$ satisfies the
following Cauchy problem for a fractional order differential
equation
\begin{align}
\label{CauchytchFBM} D_{\ast}^{\beta}v(t,x)&=
B(x,D_x)v(t,x)+\frac{\gamma
+ 1}{2}  G_{\gamma,t} A(x,D_x)  v (t,x), ~ t>0, ~ x \in \mathbb{R}^n,\\
{v}(0,x)&=\varphi(x), ~ x \in \mathbb{R}^n, \label{inCauchytchFBM}
\end{align}
where the operator $G_{\gamma,t}$ acts on the variable $t$ and is
defined by
\begin{equation} \label{presentation}
G_{\gamma,t} v(t,x)= \beta \Gamma(\gamma + 1) J^{1-\beta}_t
\mathcal{L}^{-1}_{s \to t}\big[\frac{1}{2\pi i} \int_{C-i
\infty}^{C+i\infty}
\frac{\tilde{v}(z,x)}{(s^{\beta}-z^{\beta})^{\gamma+1}}dz \big](t),
\end{equation}
with $0 < C <s,$ and $z^{\beta}=e^{\beta \mbox{Ln} (z)},$
$\mbox{Ln}(z)$ being the principal value of the complex $\ln(z).$
\end{theorem}

\textit{Proof.} Using the properties of $f_t(\tau)$ we obtain
\begin{align*}
D_{\ast,t}^{\beta}v(t,x)&= B(x,D_x) v(t,x) +
\frac{\gamma + 1}{2} A(x,D_x) \int_0^{\infty} f_t(\tau)\tau^{\gamma}
u(\tau,x)d\tau \\&= B(x,D_x) v(t,x) + \frac{\gamma + 1}{2} A(x,D_x)
G_{\gamma,t} v(t,x),
\end{align*}
where
\begin{equation} \label{gt}
G_{\gamma,t} v(t,x) = \int_0^{\infty}
f_t(\tau)\tau^{\gamma}u(\tau,x)d\tau.
\end{equation}
It follows from the definition \eqref{relation} of $v(t,x)$ that if
$\gamma =0,$ then $G_{0,t}$ is the identity operator. To show
representation \eqref{presentation} in the case $\gamma \ne 0$ we
find the Laplace transform of $G_{\gamma,t} v(t,x).$ In accordance
with the property $(d)$ of Lemma \eqref{lemproperties} we have
\[
\mathcal{L}[G_{\gamma,t} v(t,x)](s) = s^{\beta -
1}\int_0^{\infty}e^{-\tau s^{\beta}}\tau^{\gamma} u(\tau,x) d \tau =
s^{\beta-1}\mathcal{L}[\tau^{\gamma} u(\tau,x)](s^{\beta}).
\]
Obviously, if $\gamma = 0,$ then $\mathcal{L}[G_{0,t} v(t,x)](s)=
s^{\beta-1}\tilde{u}(s^{\beta},x),$ which implies $\tilde{v}(s,x) =
s^{\beta-1}\tilde{u}(s^{\beta},x).$ If $\gamma \ne 0,$ then
\begin{equation} \label{conv}
\mathcal{L}[t^{\gamma} u(t,x)](s) = \mathcal{L}[t^{\gamma}](s) \ast
\tilde{u}(s,x)=\frac{1}{2 \pi
i}\int_{c-i\infty}^{c+i\infty}\frac{\Gamma(\gamma +
1)}{(s-z)^{\gamma+1}}\tilde{u}(z,x)dz,
\end{equation}
where $\ast$ stands for the convolution of Laplace images of two
functions, and $0<c<s.$ Now using the substitution $z=e^{\beta
\mbox{Ln} (\zeta)},$ with $\mbox{Ln} (\zeta)$ the principal part of
the complex function $\ln (\zeta),$ the right hand side of
\eqref{conv} reduces to
\begin{align}
\mathcal{L}[t^{\gamma} u(t,x)](s) &= \frac{\beta}{2 \pi
i}\int_{C-i\infty}^{C+i\infty}\frac{\Gamma(\gamma +
1)}{(s-\zeta^{\beta})^{\gamma+1}}\zeta^{\beta
-1}\tilde{u}(\zeta^{\beta},x)d \zeta \notag \\& = \frac{\beta}{2 \pi
i}\int_{C-i\infty}^{C+i\infty}\frac{\Gamma(\gamma +
1)}{(s-\zeta^{\beta})^{\gamma+1}} \tilde{v}(\zeta,x)d \zeta.
\label{conv1}
\end{align}
The last equality uses the relation $\tilde{v}(\zeta,x)=
{\zeta}^{\beta-1} \tilde{u}(\zeta^{\beta},x).$ Further, replacing
$s$ by $s^{\beta}$ and taking the inverse Laplace transform in
\eqref{conv1} we obtain the desired representation
\eqref{presentation} for {the} operator $G_{\gamma,t}.$ In
accordance with part $(a)$ of Lemma \ref{lemproperties} we have
$v(0,x)=u(0,x),$ which completes the proof. \eproof

\vspace{3mm}
In the more general case when the time-change process $E_t$ is the
inverse to $W^{\mu}_t,$ the mixture of stable subordinators  with
the mixing measure $\mu,$ a representation for the FPK equation is
given in the following theorem.

\begin{theorem} \label{thdodefbm}
Let 
$u(t,x)$ be a solution to the Cauchy problem
\eqref{CauchyFBM}--\eqref{incon}. Let $f^{\mu}_t(\tau)$ be the
density function of the process inverse to $W^{\mu}_t.$ Then
$v(t,x)=\int_0^{\infty}f^{\mu}_t(\tau)u(\tau,x)d\tau$ satisfies the
following Cauchy problem for a fractional order differential
equation
\begin{align}
\label{CauchytchFBMdode} D_{\mu} v(t,x)&=
B(x,D_x)v(t,x)+\frac{\gamma
+ 1}{2}  G^{\mu}_{\gamma,t} A(x,D_x)  v (t,x), ~ t>0, ~ x \in \mathbb{R}^n,\\
v(0,x)&=\varphi(x), ~ x \in \mathbb{R}^n. \label{inCauchytchFBMdode}
\end{align}
The operator $G^{\mu}_{\gamma,t}$ acts on {the} variable $t$ and
{is} defined by
\begin{equation} \label{presentation1}
G^{\mu}_{\gamma,t} v(t,x)= \Phi_{\mu} (t)
 \ast
\mathcal{L}^{-1}_{s \to t}\big[\frac{\Gamma(\gamma+1)}{2\pi i}
\int_{C-i \infty}^{C+i\infty} \frac{ m_{\mu}(z)
\tilde{v}(z,x)}{(\rho(s)-\rho(z))^{\gamma+1}}dz \big](t),
\end{equation}
where $\ast$ denotes the usual convolution of two functions,
$0<C<s,$ $\rho(z)= \int_0^1 e^{\beta \mbox{Ln} (z)}d\mu(\beta),$
$m_{\mu}(z)=\frac{\int_0^1 \beta z^{\beta} d \mu(\beta)}{\rho(z)},$
and $\Phi_{\mu}(t)$ is defined in \eqref{phi}.
\end{theorem}

\textit{Proof.} The proof is similar to the proof of Theorem
\ref{thfbm}. We only sketch how to obtain representation
\eqref{presentation1} for the operator $G^{\mu}_{\gamma,t}
v(t,x)=\int_0^{\infty}f_t^{\mu}(\tau)\tau^{\gamma}u(\tau,x)d\tau.$
The Laplace transform of $G^{\mu}_{\gamma,t} v(t,x)$ due to part
$(d)$ of Lemma \ref{lemgen}, is
\[
\mathcal{L}_{t\to s}\big[G^{\mu}_{\gamma,t} v(t,x)\big](s) =
\frac{\rho(s)}{s} \mathcal{L}[t^{\gamma}u(t,x)](\rho(s)), s>0.
\]
Since $\mathcal{L}[\Phi_{\mu}](s)=\frac{\rho(s)}{s}, s>0,$ we have
\[
G^{\mu}_{\gamma,t} v(t,x)=\Phi_{\mu}(t) \ast \mathcal{L}_{s \to
t}^{-1}\big[\mathcal{L}[t^{\gamma}u(t,x)](\rho(s))\big](t).
\]
Further, replacing $s$ by $\rho(s)$ in \eqref{conv}, and using the
substitution $z=\rho(\zeta)=\int_0^{1}e^{\beta Ln(\zeta)}d
\mu(\beta),$ in the integral on the right side of \eqref{conv}
yields the form \eqref{presentation1}. \eproof

\vspace{3mm}

The following theorem represents the general case when the
time-change process $E_t$ is not necessarily the first hitting time
process for a stable subordinator or their mixtures.

\begin{theorem} \label{thgen}
Let $\gamma \in (-1,1).$ Let $E_t$ be a time-change {process} and
assume that $K(t,\tau)=f_{E_{_t}}(\tau)$ satisfies the condition:
$\lim_{\tau \to +0}\big[(\frac{t}{\tau})^{\gamma}K(t,\tau)\big] <
\infty$ for all $t>0.$ Let $H_t$ be an operator acting in the
variable $t$ such that
\begin{equation}
H_t K(t,\tau)=-\frac{\partial}{\partial \tau }
{\Bigl[}K(t,\tau)(\frac{t}{\tau})^{\gamma} {\Bigr]} - \delta_0(\tau)
\lim_{\tau \to
+0}{\Bigl[}(\frac{t}{\tau})^{\gamma}K(t,\tau){\Bigr]}.
\end{equation}
Then the function $v(t,x)=\int_0^{\infty}K(t,\tau)u(\tau,x)d\tau,$
where $u(t,x)$ is a solution to the Cauchy problem
\eqref{CauchyFBM}--\eqref{incon}, satisfies the equation
\begin{equation}\label{KEtch(h<0.5)}
H_t v(t,x)= t^{\gamma}\bar{G}_{-\gamma,t}B(x,D_x)v(t,x) +
\frac{\gamma+1}{2}t^{\gamma} A(x,D_x)v(t,x), ~ t>0, ~ \tau
>0,
\end{equation}
and the initial condition $v(0,x)=u(0,x).$ Here
$\bar{G}_{-\gamma,t}v(t,x)=\int_0^{\infty}K(t,\tau)\tau^{-\gamma}u(\tau,x)d\tau.$
\end{theorem}

\begin{remark}
\begin{em}
Obviously, if $\gamma \neq 0,$ then $H_t$ can not be a fractional
derivative in the sense of Caputo (or Riemann-Liouville). A
representation of $H_t$ in cases when $E_t$ is the inverse to a
stable subordinator, is given in Corollary \ref{Corr}.
\end{em}
\end{remark}

\textit{Proof.} We have
\begin{align}
H_t v(t,x)&=\int_0^{\infty} H_t K(t,\tau) u(\tau,x) d \tau
\notag\\
&= - \int_0^{\infty} \Big\{\frac{\partial}{\partial \tau} {\Bigl[}
K(t,\tau)(\frac{t}{\tau})^{\gamma} {\Bigr]}+ \delta_0(\tau)
\lim_{\tau \to +0}{\Bigl[}(\frac{t}{\tau})^{\gamma}K(t,\tau) {\Bigr]}\Big\} u(\tau,x)d\tau \notag\\
&= -t^{\gamma} \lim_{\tau \to \infty} [K(t,\tau)
\tau^{-\gamma}u(\tau,x)] + t^{\gamma} \lim_{\tau \to 0+} [K(t,\tau)
\tau^{-\gamma}u(\tau,x)] \label{zeros}\\
&+\int_0^{\infty} K(t,\tau)\big(\frac{t}{\tau}\big)^{\gamma}
\frac{\partial u(\tau,x)}{\partial \tau} d \tau - \lim_{\tau \to
+0}{\Bigl[}(\frac{t}{\tau})^{\gamma}K(t,\tau){\Bigr]}u(0,x). \notag
\end{align}
Obviously, the first term on the right of \eqref{zeros} is equal to
zero, since for each fixed $t>0$ function $K(t,\tau)$ is bounded
when $\tau \to \infty$ and $u(\tau,x)$ decays at infinity. The sum
of the second and last terms, which exist by the hypothesis of the
theorem, also equals zero. Now taking {equation} \eqref{CauchyFBM}
 into account,
\begin{align*}
H_t v(t,x) &= t^{\gamma} B(x,D_x)
\int_0^{\infty}K(t,\tau){\tau}^{-\gamma}u(\tau,x)d\tau+
\frac{\gamma+1}{2} t^{\gamma} A(x,D_x) v(t,x).
\end{align*}

Further, since $E_0=0$ it follows that
\[
\hspace{2cm} \lim_{t \to 0} v(t,x)=\int_0^{\infty}
\delta_0(\tau)u(\tau,x)d\tau=u(0,x). \ \hspace{5cm} \eproof
\]

Let $\Pi_{\gamma}$ denote the operator of multiplication by
$t^{\gamma},$ i.e. $\Pi_{\gamma} h(t)=t^{\gamma}h(t), ~ h \in
C(0,\infty).$ Applying Theorem \ref{thgen} to the case
$K(t,\tau)=f_t(\tau)$ in conjunction with Theorem \ref{thfbm}, we
obtain the following corollary.

\begin{corollary} \label{Corr} Let $\gamma \le 0$ and
$K(t,\tau)=f_t(\tau),$ where $f_t(\tau)$ is defined in \eqref{relation2}. Then
\begin{enumerate}
\item[(i)]
$G_{-\gamma,t}=G_{\gamma,t}^{-1};$
\item[(ii)]
$H_t = \Pi_{\gamma} G_{-\gamma,t} D_{\ast}^{\beta}.$
\end{enumerate}
\end{corollary}

This Corollary yields an equivalent form for FPK equation
\eqref{CauchytchFBM} in the case when $E_t$ is the inverse to the
stable subordinator with index $\beta$ and $\gamma \le 0:$
\begin{equation}\label{equiv}
H_t v(t,x)= t^{\gamma}{G}_{-\gamma,t}B(x,D_x)v(t,x) +
\frac{\gamma+1}{2}t^{\gamma} A(x,D_x)v(t,x),
\end{equation}
with $H_t$  {as} in Corollary \ref{Corr}.

Suppose the operator in the drift term $B(x,D_x)=0.$ Then equation
\eqref{equiv} takes the form
\begin{equation}\label{equiv-0}
H_t v(t,x)= 
\frac{\gamma+1}{2}t^{\gamma} A(x,D_x)v(t,x). 
\end{equation}
Consequently, given a FPK equation associated to an SDE driven by a
non-time-changed fBM, the FPK equation for the analogous SDE driven
by the time-changed fBM cannot be of {the} form: retain the right
hand side and change the left hand side to a fractional derivative.
Moreover, if a fractional derivative is desired on the left hand
side in the time-changed case, then the right hand side must be a
different operator from that in the non-time-changed case.

Notice that FPK equation \eqref{equiv} is valid for $\gamma \in
(0,1)$ as well. Indeed, part $(ii)$ of Corollary \ref{Corr} can be
rewritten in the form $G_{\gamma,t}=G_{-\gamma,t}^{-1}$ for $\gamma
>0.$ For $\gamma <0$ part $(ii)$ of Corollary \ref{Corr} also implies
$(G_{\gamma,t}^{-1})^{-1}=G^{-1}_{-\gamma,t}=G_{\gamma,t}.$ Now
applying operators $G_{-\gamma,t}$ and $\Pi_{\gamma}$ consecutively
to both sides of \eqref{CauchytchFBM} we obtain \eqref{equiv} for
all $\gamma \in (-1,1).$

Analogously, the FPK equation obtained in Theorem \ref{thdodefbm}
with the mixing measure $\mu$ can be represented in its equivalent
form as
\begin{equation}\label{equiv-2}
H^{\mu}_t v(t,x)= t^{\gamma}{G}^{\mu}_{-\gamma,t}B(x,D_x)v(t,x) +
\frac{\gamma+1}{2}t^{\gamma} A(x,D_x)v(t,x), ~ t>0, ~ \tau
>0,
\end{equation}
where $H^{\mu}_t = \Pi_{\gamma} G^{\mu}_{\gamma, t} D_{\mu}.$ We
leave verification of the details to the reader.

The equivalence of equations \eqref{CauchytchFBM} and
{\eqref{equiv}} and the equivalence of equations
\eqref{CauchytchFBMdode} and \eqref{equiv-2} are obtained by means
of Theorem \ref{thgen}. This fact can also be
established with the help of the semigroup property 
of the family of
operators $\{G_{\gamma}, -1 < \gamma <1\}:$ 
\begin{equation} \label{gt1}
G_{\gamma}g(t) = \int_0^{\infty}f_t(\tau)\tau^{\gamma}h(\tau)d\tau =
\mathcal{F}_{\gamma}h(t),
\end{equation}
where $h \in C^{\infty}(0,\infty)$ is a non-negative function
rapidly decreasing at infinity. Denote the class of such functions
by $U.$ Functions $g$ and $h$ in \eqref{gt1} are connected through
the relation $g(t)=\int_0^{\infty}f_t(\tau)h(\tau)d\tau
=\mathcal{F}h(t).$ It follows from the behaviour of $f_t(\tau)$ as a
function of $t,$ that $g \in C^{\infty}(0,\infty),$ but not
necessarily integrable. On the other hand, obviously, operator
$\mathcal{F}$ is bounded, $\|\mathcal{F}h\| \le \|h\|$ in the
$\sup$-norm, and one-to-one due to positivity of $f_t(\tau).$
Therefore, the inverse $\mathcal{F}^{-1}: \mathcal{F}U \to U$
exists. Let a tempered distribution $H(t,\tau)$ with $supp \, H
\subset \mathbb{R}_+^2$ be such that
$\mathcal{F}^{-1}g(t)=\int_{0}^{\infty}H(t,\tau)g(\tau)d\tau.$ Since
$f_t(\tau)\in \mathcal{F}U$ as a function of $t$ for each $\tau >
0,$ for an arbitrary $h \in U$ one has
\begin{align*}
h(t) = \mathcal{F}^{-1} \mathcal{F} h(t) &= \int_{0}^{\infty} H(t,s)
\big(\int_0^{\infty} f_{s}(\tau)h(\tau)d\tau\big) ds\\
&=\int_{0}^{\infty} h(\tau) \big(\int_0^{\infty} H(t,s)f_{s}(\tau)ds
\big) d\tau \\&= <\int_0^{\infty} H(t,s)f_{s}(\tau)ds ,h>_{\tau}.
\end{align*}
We write this relation between $H(t,\tau)$ and $f_t(\tau)$ in the
form
\begin{equation}\label{01}
\int_0^{\infty}H(t,s)f_s(\tau)ds = \delta_t{(\tau)}.
\end{equation}

\begin{proposition} \label{semigroup}
Let $-1<\gamma <1,$ $-1 < \alpha <1,$ and $-1 <\gamma + \alpha <1.$
Then $G_{\gamma} \circ G_{\alpha}= G_{\gamma + \alpha}.$
\end{proposition}

\textit{Proof.} The proof uses the following two relations:
\begin{enumerate}
\item[(1)]
$G_{\gamma}g(t)=\int_0^{\infty} \mathcal{F}_{\gamma,t}H(t,s)g(s)ds, ~ \gamma \in (-1,1);$ 
\item[(2)]
$\int_0^{\infty}\mathcal{F}_{\gamma,t}H(t,s)
\mathcal{F}_{{\alpha,s}}H(s,\tau)ds =\mathcal{F}_{{\gamma+\alpha,
t}}H(t,\tau),$ with $-1<\gamma, \alpha <1,$ and $-1 <\gamma + \alpha
<1.$
\end{enumerate}
Indeed, using \eqref{gt1} and changing the order of integration, we
obtain the first relation
\begin{align}
G_{\gamma} g(t)&= \int_0^{\infty}f_t(\tau)\tau^{\gamma}\big(
\int_0^{\infty}H(\tau,s)g(s)ds \big)d\tau \notag \\
&=\int_0^{\infty}g(s)\big(\int_0^{\infty}f_t(\tau)H(\tau,s)\tau^{\gamma}d\tau
\big)ds \notag \\
&=\int_0^{\infty}\mathcal{F}_{\gamma,t}H(t,s)g(s)ds. \label{2nd}
\end{align}
It is readily seen that the internal integral in the second line of
\eqref{2nd} is meaningful, since $f_{t}(\tau)$ is a function of
exponential decay when $\tau \to \infty,$ which follows from
\eqref{atzero}. Further, in order to show the second relation, we
have
\begin{align*} \label{03rd}
\int_0^{\infty}\mathcal{F}_{\gamma,t}H(t,s)&
\mathcal{F}_{{\alpha,s}}H(s,\tau)ds
 = \int_0^{\infty} \big(
\int_0^{\infty}f_t(p)H(p,s)p^{\gamma}dp\big)
\big( \int_0^{\infty}f_s(q)H(q,\tau)q^{\alpha}dq\big)ds\\
&=\int_0^{\infty}
\int_0^{\infty}f_t(p)H(q,\tau)p^{\gamma}q^{\alpha}\big(\int_0^{\infty}H(p,s)f_s(q)ds\big)dpdq.
\end{align*}
Due to \eqref{01}, this equals
\begin{align*}
\int_0^{\infty} f_t(p)p^{\gamma} \big(\int_0^{\infty}
H(q,\tau)q^{\alpha} \delta_p(q) dq \big)dp &=\int_0^{\infty}
H(p,\tau) p^{\alpha} f_t(p) p^{\gamma} dp
\\&= \mathcal{F}_{\gamma +
\alpha,t}H(t,\tau).
\end{align*}
Now we are ready to prove the claimed semigroup property. Making use
of the two proved relations,
\begin{align*}
(G_{\gamma}\circ
G_{\alpha})g(t)&=G_{\gamma}\big[G_{\alpha}g(t)\big]=
G_{\gamma}\big[\int_0^{\infty}\mathcal{F}_{{\alpha,t}}H(t,s)g(s)ds\big]\\
&=\int_0^{\infty}\mathcal{F}_{\gamma,t}H(t,s)\big[\int_0^{\infty}\mathcal{F}_{\alpha,s}H(s,\tau)g(\tau)d\tau\big]ds\\
&=\int_0^{\infty}g(\tau)\int_0^{\infty}\mathcal{F}_{\gamma,t}H(t,s)\mathcal{F}_{\alpha,s}H(s,\tau)dsd\tau\\
&=\int_0^{\infty}\mathcal{F}_{\gamma+\alpha,t}H(t,\tau)g(\tau)d\tau=G_{\gamma+\alpha}g(t). \ \ \eproof
\end{align*}

\begin{remark}
\begin{em} \mbox{ }
\begin{enumerate}
\item Proposition \ref{semigroup} immediately implies that
$G_{\gamma}^{-1}=G_{-\gamma}$ for arbitrary $\gamma \in (-1,1).$
Indeed, $G_{\gamma}\circ G_{-\gamma}=G_0=I,$ as well as
$G_{-\gamma}\circ G_{\gamma}=I,$ where $I$ is the identity operator.
Thus, the statement in Corollary \ref{Corr} is valid for all $\gamma
\in (-1,1).$
\item Proposition \ref{semigroup} remains valid for the family $\{G^{\mu}_{\gamma}, -1 < \gamma
<1\},$ as well.
\item The method used in this paper allows extension of results of Theorems
\ref{thfbm}--\ref{thgen} to the case of SDEs driven by time-changed
linear fractional 
{L\'evy stable} motions. See \cite{MNX} for CTRW limits of
correlated random variables, whose limiting processes are
time-changed fractional {Brownian}, or linear fractional 
{L\'evy stable} motions.
\item The formula $v(t,x)=\mathcal{F}u(t,x)$ for a solution of FPK equations
associated with time-changed fBM, provides a useful tool for
analysis of properties of a solution to initial value problems
\eqref{CauchytchFBM}--\eqref{inCauchytchFBM} and
\eqref{CauchytchFBMdode}--\eqref{inCauchytchFBMdode}, as well as to
the Cauchy problem for equation \eqref{KEtch(h<0.5)}.
\item
It is not necessary for the dependence of coefficients in
\eqref{KEfbm1} on $t$ to be of the form
 $t^{\gamma}.$ This function can be replaced by
$[\nu(t)]^{\gamma},$ where $\nu(t)$ is a continuous function defined
on $[0,\infty);$ however, the results essentially depend on the
behavior of $\nu(t)$ near zero and infinity.
\end{enumerate}
\end{em}
\end{remark}

\end{document}